# A Real-Time Detecting Algorithm for Tracking Community Structure of Dynamic Networks

Jiaxing Shang*[†], Lianchen Liu*[†], Feng Xie[†], Zhen Chen[†], Jiajia Miao[‡], Xuelin Fang[§], Cheng Wu*[†]
[†]Department of Automation, Tsinghua University, Beijing, 100084, China
*National CIMS Engineering Center, Tsinghua University, Beijing, 100084, China
[‡]Institute of Command Automation, PLA University of Science and Technology, Najing, 210007, China
[§]Institute of Computer Science, National University of Defense Technology, Changsha, 410073, China
{xief10, shangjx06}@mails.tsinghua.edu.cn, {liulianchen, zhenchen}@tsinghua.edu.cn

## ABSTRACT
In this paper a simple but efficient real-time detecting algorithm is proposed for tracking community structure of dynamic networks. Community structure is intuitively characterized as divisions of network nodes into subgroups, within which nodes are densely connected while between which they are sparsely connected. To evaluate the quality of community structure of a network, a metric called modularity is proposed and many algorithms are developed on optimizing it. However, most of the modularity based algorithms deal with static networks and cannot be performed frequently, due to their high computing complexity. In order to track the community structure of dynamic networks in a fine-grained way, we propose a modularity based algorithm that is incremental and has very low computing complexity. In our algorithm we adopt a *two-step* approach. Firstly we apply the Blondel *et al*'s algorithm for detecting static communities to obtain an initial community structure. Then, apply our incremental updating strategies to track the dynamic communities.

The performance of our algorithm is measured in terms of the modularity. We test the algorithm on tracking community structure of Enron Email and three other real world datasets. The experimental results show that our algorithm can keep track of community structure in time and outperform the well known CNM algorithm in terms of modularity.

## Categories and Subject Descriptors
H.2.8 [Database Applications]: Data mining.

## 1. INTRODUCTION
Many systems can be represented as networks or graphs [1, 2, 3], in which nodes represent the individuals while edges represent the relationship or interactions between nodes. Examples include social networks [4], citation networks [5, 6, 7], biological networks [8, 9, 18], mobile phone networks [10, 11, 25], terrorism networks [12, 13, 14], etc. One of the most important problems in network analysis is the identification of community structure, the division of network nodes into subgroups, within which nodes are densely connected while between which they are sparsely connected [15]. The identification of communities often reveals deeper properties of networks and has many important applications, such as detecting criminal organizations, finding common hobbies shared by people, etc. To combat the problem of finding community structure, many algorithms have been proposed recently. Girvan and Newman [16] proposed a divisive algorithm that uses edge betweenness as a metric to identify the boundaries of communities. In this algorithm, Newman proposed *modularity (Q)* as a metric to evaluate the quality of a community structure and successfully applied it to a variety of networks. However, as noted in [16], the algorithm is time-consuming, which restricts its application to only small networks.

After Newman proposed his modularity-based algorithm, a number of faster algorithms have been proposed [15, 18, 19, 20, 21]. Since finding a global optimization $Q$ is a NP-hard problem [16, 22], most of these algorithms take advantage of heuristic strategies. To the best of our knowledge, the fastest modularity optimization based algorithm is proposed by Blondel *et al* [20] (Blondel, Guillaume, Lambiotte and Lefebvre, in the rest of the paper, we will call this algorithm "the BGL algorithm"). Experimental results show that this algorithm outperforms all other state-of-art modularity based algorithms in terms of computation time (It reveals communities of a web network of 118 million nodes and more than one billion links within 152 minutes). Moreover, the quality of the community structure detected is quite good, as measured by the modularity.

The modularity optimization based algorithms mentioned above have been applied successfully in many real world networks and some of them run quite fast [15, 20]. However, most of them just focus on the representation of graphs as static networks and only give some snapshots of networks. When the networks change, these algorithms have to be re-performed to get the community structure, making it impossible for them to track the community structure on a fine-grained level. Real world networks are usually dynamic and some of them change frequently. For some networks their community structure needs to be tracked timely. For example, in the terrorism networks the criminals usually do not communicate with each other until they are to commit the crime. We have to be able to track the community structure so as to reveal the criminal organizations and avoid terrorist incident before it is too late.

In order to keep track of the community structure of a network on a fine-grained level, our algorithm should have very low computing complexity in updating the community structure when network changes. To combat this problem we model the change of networks as sequential increase of edges (this is why we call our algorithm incremental). We classify increased edges into four types: *i)* inner community edge, *ii)* cross community edge, *iii)* half-new edge, and *iv)* new edge. In our *two-step* approach, we firstly apply the BGL algorithm to generate an initial community structure of networks. Then we apply different strategies to update the community structure according to the edges' type. Which strategy to apply follows a basic principle: the strategy should be able to increase the modularity of the community structure, if not, it should make the lost in modularity as low as possible. The running time for updating the community structure when increasing an edge varies from $O(1)$ to $O(S)$, where $S$ is the size of the community to be updated. From the experimental results we found that the incremental algorithm runs at the computing complexity of $O(1)$ in most of the time, which guarantees the high efficiency of our algorithm. The quality of our algorithm is evaluated by the modularity. We test the performance on four real-world datasets, including email communication network, paper citation network, web voting network, etc. The performance is compared with that of the BGL [20] and CNM [15] algorithm. Experimental results show that our algorithm outperforms both BGL and CNM in computing time and outperforms CNM in terms of modularity.

The rest of the paper is organized as follows: Section 2 gives some related work in dynamic networks. Section 3 introduces the modularity and BGL algorithm. Section 4 presents our incremental algorithm for tracking community structure of dynamic networks; Section 5 gives experimental results to validate the effectiveness and the high computational efficiency of our algorithm. Section 6 draws the conclusion.

## 2. RELATED WORK

Recently significant attention is attracted to the literature of community detection in dynamic networks. In [17] the authors from Yahoo! studied the evolution of community structure within two large online social networks: Flicker and Yahoo! 360. Since they have access to the entire lifetime of the two networks, they are able to study their dynamic properties in depth. By researching the evolution of edge density, community diameter, delay in reciprocity, degree and node distribution in the two dynamic networks, the authors reveal an interesting unexpected growing rule: rapid growth, decline, and then slow but steady growth.

C. Tantipathananandh and T. Berger-Wolf [23] model the identification of communities in dynamic networks as a combinatorial optimization problem (a coloring problem) based on some basic assumptions about the behavior of individuals. The optimization problem is shown to be NP-hard, so the authors present approximation algorithms using dynamic programming and greedy heuristics to solve it. In [27] Tantipathananandh and Berger-Wolf extend their previous model to arbitrary dynamic networks and approximately solve the optimization problem using semidefinite programming relaxation and a rounding heuristic.

Greene etc. [24] use an event model to describe the evolution of dynamic networks where the life-cycle of each community is characterized by a series of significant events. These events include *birth*, *death*, *merging*, *splitting*, *expansion* and *contraction* of dynamic communities. In this paper a dynamic network is represented as a set of *time step* graphs. Firstly *step communities* are identified at individual time steps. The step communities can be found using arbitrary community detection algorithms. Then the authors present a simple algorithm to match the communities found at consecutive time steps. The algorithm is evaluated on both synthetic dynamic networks and a real-world mobile operator call network.

## 3. PRELIMINARIES

The modularity is proposed by Newman [16] and used to evaluate the quality of network's community structure. Let $G=(V,E)$ be an undirected weighted graph [26], where $V$ is the set of nodes and $E$ is the set of edges between nodes. Then we can use an adjacency matrix $A$ to describe the graph:

$$A_{ij} = \begin{cases} w_{ij} & \text{if nodes i and j are connected,} \\ 0 & \text{otherwise} \end{cases} \quad (1)$$

where $w_{ij}$ is the weight of edge connecting nodes $i$ and $j$. In [16] Newman gives the definition of *modularity* as follows:

$$Q = \frac{1}{2m} \sum_{i,j \in V} \left[ A_{ij} - \frac{k_i k_j}{2m} \right] \delta(c_i, c_j) \quad (2)$$

where $c_i$ is the community that node $i$ belongs to, $m = \Sigma_{i,j \in V} A_{ij} / 2$ is the sum of weights of edges in the graph, $k_i = \Sigma_{j \in V} A_{ij}$ is the weighted degree of node $i$, the $\delta$ function $\delta(u,v)$ is 1 if $u=v$ and 0 otherwise. Another expression of $Q$ is:

$$Q = \sum_{c \in C} \left( e_{cc} - a_c^2 \right) \quad (3)$$

where $e_{vw} = \frac{1}{2m} \sum_{i,j \in V} A_{ij} \delta(c_i, v) \delta(c_j, w)$ is the fraction of weights of edges that connecting nodes in community $v$ and nodes in community $w$, $a_v = \frac{1}{2m} \sum_{i \in V} k_i \delta(c_i, v)$ is the fraction of weights of edges that are attached to nodes in community $v$.

The expression of $Q$ consists of two parts. Intuitively, it measures the fraction of edges within communities minus the expected fraction of edges in a network with the same community divisions but random connections between nodes. Then the task of community detection turns to the task of finding a division corresponding to high value of $Q$.

Unfortunately, finding the exact division of a network corresponding to the highest modularity is a problem that is computationally hard [22]. So approximation algorithms are necessary when dealing with large networks. To the best of our knowledge, the fastest approximation algorithm for optimizing modularity on large networks is proposed by Blondel *et al* [20] (the BGL algorithm), which applies a two-phase iteration approach to generate hierarchical community structure. The idea of this efficient algorithm is motivated by the fact that the gain in modularity $\Delta Q$ obtained by moving an isolated node $i$ to a community $c \in C$ can easily be computed by

$$\Delta Q = \left[ \frac{\Sigma_{in}^c + 2k_{i,in}}{2m} - \left( \frac{\Sigma_{tot}^c + k_i}{2m} \right)^2 \right] - \left[ \frac{\Sigma_{in}^c}{2m} - \left( \frac{\Sigma_{tot}^c}{2m} \right)^2 - \left( \frac{k_i}{2m} \right)^2 \right]$$

$$= \frac{1}{2m} \left( 2k_{i,in} - \frac{\Sigma_{tot}^c k_i}{m} \right) \quad (4)$$

where $\Sigma_{in}^c = \sum_{i,j \in V} A_{ij} \delta(c_i, c) \delta(c_j, c)$ is the sum of the weights of the edges inside $c$, $\Sigma_{tot}^c = \sum_{i \in V} k_i \delta(c_i, c)$ is the sum of the weights of the edges incident to nodes in $c$, $k_i$ is the sum of the weights of the edges incident to node $i$, $k_{i,in}$ is the sum of the weights of the edges from $i$ to nodes in $c$ and $m$ is the sum of the weights of all the edges in the network. There is another similar expression used to evaluate the change of modularity when node $i$ is removed from its community. Then we can evaluate the change of modularity when a node changes its community by removing it from its current community and then by moving it into the new community.

This algorithm applies a two-phase approach. Initially, each node is placed in a separate community. In the first phase, for each node we check if moving the node from its current community to any community to which a neighbor belongs would yield an increase in modularity. If so, the node is moved to the neighboring community that gives the highest gain in modularity. The process is continued until all nodes are processed. Then in the second phase we convert each "updated" community to a single node (called community-node) in a new network, with edges between community-nodes where there are edges between nodes in the communities of the original network. The weights of the new edges are obtained by summing over all previous weights. For each new node, there is an edge called self-loop, whose weight is equal to the sum of the weights of previous edges inside the community. Then, return to the first phase and run the next round of iteration. The algorithm will terminate when the modularity gain between two rounds of iteration is lower than a given threshold. The final results correspond to a hierarchical community structure.

The algorithm runs quickly, taking just 150 minutes to run on a network of 118 million nodes and 1 billion edges. Moreover, the detected community structures tend to have high modularity than other algorithms mentioned in [20].

## 4. INCREMENTAL ALGORITHM

The BGL algorithm performs quite well on static networks. However, when the network changes, the algorithm has to be reperformed to track the community structure. This is obviously impossible if the networks change frequently or when they become very large. Tantipathananandh, Greene, etc. propose algorithms to find communities in dynamic networks. However, there is a common shortcoming in these algorithms: they are unable to track the communities on a fine-grained level. Some of them [23, 27] can handle only small networks, due to the high computing complexity in solving optimization problems.

### 4.1 Type of Edges

In order to tackle these problems, we propose an incremental algorithm with low computing complexity. We view the evolution of networks as the observation of interactions between individuals. For example, Bob and Alice are closer together because we observe that they interact more often with each other, a new member joined in the community because we observe that he or she interacts with at least one of the members in the community. This is quite a different perspective from that of state-of-art studies. Then we can model the change of networks as sequential observation of interactions, which can be represented as sequential increase of edges (this is why we call our algorithm incremental). For each increased edge $l(i, j, w_{ij})$, where $i$ is the source node, $j$ is the target node and $w_{ij}$ is the weight, it could be assigned to one of the four types:

1. Inner community edge: the two nodes incident to the edge already exist and belong to the same community.

2. Cross community edge: the two nodes incident to the edge already exist and belong to different communities.

3. Half-new edge: one of the nodes incident to the edge is new.

4. New edge: both of the nodes incident to the edge are new.

Figure 1 shows the four types of increased edges and the change it brings to the network. The original network graph is divided into two communities {1, 2} and {3, 4, 5} according to the BGL algorithm, as shown on the left side. After the four types of edges are added, the community structure is affected in different ways, as shown on the right side.

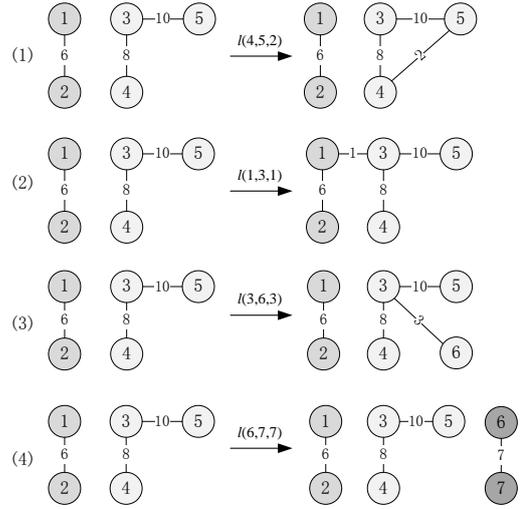

**Figure 1. The types of edges**

### 4.2 Algorithm

Then we take different operations to update the community structure according to the edges' type. These operations include: 1. keeping the community structure unchanged; 2. combining two communities into one; 3. assigning nodes to an existing community; 4. creating a new community with new nodes. We see from these operations that operation 2 reduces the number of communities while operation 4 increases it. Operation 1 actually does nothing on the community structure. Which operation to be taken follows a basic principle: the operation should be able to increase the modularity of the community structure, if not, it should make the lost in modularity as low as possible. In Eq. (3), by writing $e_{cc} = \Sigma_{in}^c/2m$ and $a_c = \Sigma_{tot}^c/2m$, We give another expression of Eq. (3) as follows:

$$Q = \frac{1}{2m}\sum_{c\in C}\left(\Sigma_{in}^c - \frac{\Sigma_{tot}^{c\ 2}}{2m}\right) \qquad (5)$$

It will be used to decide the operation to be taken.

Now we illustrate in detail how to choose operations to update the community structure when an increased edge $l(i, j, w_{ij})$ comes.

1. If the increased edge $l(i, j, w_{ij})$ is an inner community edge, as illustrated in Figure 1, we see that it increases the inner connections of the community and keeps the inter-community connections unchanged, which coincides with the basic principle of the modularity, so operation 1 will be taken, in this case, the community structure will keep unchanged.

2. If the increased edge $l(i, j, w_{ij})$ is a cross community edge, which means nodes $i$ and $j$ belong to two different communities, suppose they are $c_i$ and $c_j$ respectively. Then two candidate operations will be taken: operation 1 or operation 2. The former keeps the community structure unchanged, while the latter combines communities $c_i$ and $c_j$ into one. Our purpose is to increase the modularity or make the loss of it as low as possible. So we compare the modularity gain brought by these two operations. If the community structure of the network keeps unchanged, the new modularity value will be:

$$Q_1 = \frac{1}{2m+2w_{ij}} \left\{ \sum_{c \in C}^{c \neq c_i, c_j} \left( \Sigma_{in}^c - \frac{\Sigma_{tot}^{c~2}}{2m+2w_{ij}} \right) + \left( \Sigma_{in}^{c_i} - \frac{(\Sigma_{tot}^{c_i} + w_{ij})^2}{2m+2w_{ij}} \right) + \left( \Sigma_{in}^{c_j} - \frac{(\Sigma_{tot}^{c_j} + w_{ij})^2}{2m+2w_{ij}} \right) \right\} \quad (6)$$

If the two communities $c_i$ and $c_j$ are combined into one, the new modularity value can be formulated as:

$$Q_2 = \frac{1}{2m+2w_{ij}} \left\{ \sum_{c \in C}^{c \neq c_i, c_j} \left( \Sigma_{in}^c - \frac{\Sigma_{tot}^{c~2}}{2m+2w_{ij}} \right) + \left( \Sigma_{in}^{c_i} + \Sigma_{in}^{c_j} + 2w_{ij} - \frac{(\Sigma_{tot}^{c_i} + \Sigma_{tot}^{c_j} + 2w_{ij})^2}{2m+2w_{ij}} \right) \right\} \quad (7)$$

Suppose the modularity of the community structure before any operations are taken is $Q$, then the modularity gain brought by the two operations will be $\Delta Q_1 = Q_1 - Q$ and $\Delta Q_2 = Q_2 - Q$ respectively. By comparing them, we get:

$$\Delta Q_2 - \Delta Q_1 = Q_2 - Q_1 = \frac{1}{2m+2w_{ij}} \left( 2w_{ij} - \frac{2(\Sigma_{tot}^{c_i} + w_{ij})(\Sigma_{tot}^{c_j} + w_{ij})}{2m+2w_{ij}} \right) \quad (8)$$

When $w_{ij}(2m+2w_{ij}) > 2(\Sigma_{tot}^{c_i} + w_{ij})(\Sigma_{tot}^{c_j} + w_{ij})$, namely, $\Delta Q_2 > \Delta Q_1$, we take operation 2, else operation 1.

3. If the increased edge $l(i, j, w_{ij})$ is a half-new edge, which means that one of the nodes incident to $l$ (assume $i$) already exists in the network while the other (node $j$) is new (there are no edges between $j$ and any of the other nodes before $l$ is increased). Two candidate operations will be taken: operation 3 or operation 4. The former assigns $j$ to the community that $i$ belongs to (assume $c_i$), while the latter creates a new community $c_j$ and assign $j$ to it. Similarly, we compare the modularity gain brought by the two operations. If $j$ is assigned to community $c_i$, the new modularity value will be:

$$Q_1 = \frac{1}{2m+2w_{ij}} \left\{ \sum_{c \in C}^{c \neq c_i} \left( \Sigma_{in}^c - \frac{\Sigma_{tot}^{c~2}}{2m+2w_{ij}} \right) + \left( \Sigma_{in}^{c_i} + 2w_{ij} - \frac{(\Sigma_{tot}^{c_i} + 2w_{ij})^2}{2m+2w_{ij}} \right) \right\} \quad (9)$$

If a new community $c_j$ is created for $j$, the new modularity value will be:

$$Q_2 = \frac{1}{2m+2w_{ij}} \left\{ \sum_{c \in C}^{c \neq c_i} \left( \Sigma_{in}^c - \frac{\Sigma_{tot}^{c~2}}{2m+2w_{ij}} \right) + \left( \Sigma_{in}^{c_i} - \frac{(\Sigma_{tot}^{c_i} + w_{ij})^2}{2m+2w_{ij}} \right) + \left( 0 - \frac{w_{ij}^2}{2m+2w_{ij}} \right) \right\} \quad (10)$$

Comparing the modularity gain brought by the two operations, we get:

$$\Delta Q_2 - \Delta Q_1 = -\frac{2w(2m - \Sigma_{tot}^{c_i}) + 2w_{ij}^2}{(2m+2w_{ij})^2} \quad (11)$$

Since $2m - \Sigma_{tot}^i \geq 0$ is identically true, Eq. (11) will always be smaller than zero, which means operation 3 will bring more gain or less loss in modularity than operation 4, so operation 3 will be taken to update the community structure.

4. If the increased edge $l(i, j, w_{ij})$ is a new edge, which means that both of the nodes incident to $l$ are new (there are no edges among $i, j$ and any of the other nodes of the network before $l$ is increased). Two candidate operations will be taken: operation 3 or operation 4. The former assigns $i$ and $j$ to an existing community (assume $c_k$), while the latter creates a new community $c_i$ and put both $i$ and $j$ into it. Again we compare the modularity gain brought by the two operations. If we assign $i, j$ to an existing community $c_k$, the new modularity value will be:

$$Q_1 = \frac{1}{2m+2w} \left\{ \sum_{c \in C}^{c \neq c_k} \left( \Sigma_{in}^c - \frac{\Sigma_{tot}^{c~2}}{2m+2w} \right) + \left( \Sigma_{in}^{c_k} + 2w - \frac{(\Sigma_{tot}^{c_k} + 2w)^2}{2m+2w} \right) \right\} \quad (12)$$

If we create a new community $c_i$ and put both $i$ and $j$ into it, the new modularity value will be:

$$Q_1 = \frac{1}{2m+2w} \left\{ \sum_{c \in C}^{c \neq c_k} \left( \Sigma_{in}^c - \frac{\Sigma_{tot}^{c~2}}{2m+2w} \right) + \left( \Sigma_{in}^{c_k} - \frac{\Sigma_{tot}^{c_k~2}}{2m+2w} \right) + \left( 2w - \frac{(2w)^2}{2m+2w} \right) \right\} \quad (13)$$

Comparing the modularity gain brought by the two operations, we get:

$$\Delta Q_2 - \Delta Q_1 = \frac{4w\Sigma_{tot}^{c_k}}{(2m+2w)^2} \quad (14)$$

where $\Delta Q_2 > \Delta Q_1$ will be identically true, so operation 4 will be taken.

The relationship between different types of edges and the corresponding operations adopted on them are summarized in Table 1, in which row items represent the operations while column items represent the different types of edges. In this table, "ICE" is short for "Inner community edge", "CCE" for "Cross community edge", "HNE" for "Half-new edge" and "NE" for "New edge". "Y" means the operation will be taken on the corresponding type of edge, "P" means the operation is possible to be taken on the edge and "N" means the operation will not be taken on it.

**Table 1. Relationship between different types of edges and operations**

|       | ICE | CCE | HNE | NE |
|-------|-----|-----|-----|----|
| Opt 1 | Y   | P   | N   | N  |
| Opt 2 | N   | P   | N   | N  |
| Opt 3 | N   | N   | Y   | N  |
| Opt 4 | N   | N   | N   | Y  |

From this table we see that if the increased edge is a cross community edge, operation 1 and operation 2 are possible to be taken. The algorithm is described in Algorithm 1.

---

**Initialize**: Run the GBL algorithm to generate an initial community structure
**for** $i := 1$ **to** Number of increased edges **do**
    **switch** Typeof(edge$_i$)
    **case**: ICE
        Keep community structure unchanged.
    **case**: CCE
        **if** (deltaQ(opt 2) > deltaQ(opt 1)) **then**
            Combine communities with opt 2.
        **else** Keep community structure unchanged.
        **end**
    **case**: HNE
        Update community structure with opt 3.
    **case**: NE
        Update community structure with opt 4.
**end**

**Algorithm 1**: The incremental algorithm

## 4.3 Deeper Inspection of the Algorithm

In this part we will take a deep look at the incremental algorithm. The incremental algorithm is designed for tracking community structure of dynamic networks of which the change is incremental and frequent. Suppose an extreme example: the initial network contains no nodes or edges ($G(V,E), V = \phi, E = \phi$) and the community structure is incrementally generated with our algorithm. The final network is shown in Figure 2(1), which consists of 6 nodes and 7 edges. We firstly consider the sequence of increased edges as {(1, 2, 13), (1, 3, 8), (2, 3, 6), (4, 5, 12), (4, 6, 9), (5, 6, 5), (3, 4, 2)}. When the first edge (1, 2, 13) is increased, from the definition of the types of edges, we see it is a new edge, so operation 4 will be taken, in this case, a new community $c_1$ is created and nodes 1 and 2 are assigned to it. Then edge (1, 3, 8) comes, since node 1 already exists in the network now, operation 3 will be taken and vertex 3 is assigned to $c_1$, the community that vertex 1 belongs to. The next increased edge is (2, 3, 6), which is an inner community edge, operation 1 will be taken and each node keeps its community affiliation unchanged. Then edge (4, 5, 12) is increased and a new community $c_4$ is created. This process keeps on until all the edges are increased. Finally we get the community structure of the network, which consists of two communities. The nodes of the two communities are colored, as shown in Figure 2(2). Let us consider another circumstance and make the sequence of increased edges as {(1, 2, 13), (1, 3, 8), (2, 3, 6), (3, 4, 2), (4, 5, 12), (4, 6, 9), (5, 6, 5)}. After the execution of the incremental algorithm, we will get a community structure consisting of only one community, as shown in Figure 2(3).

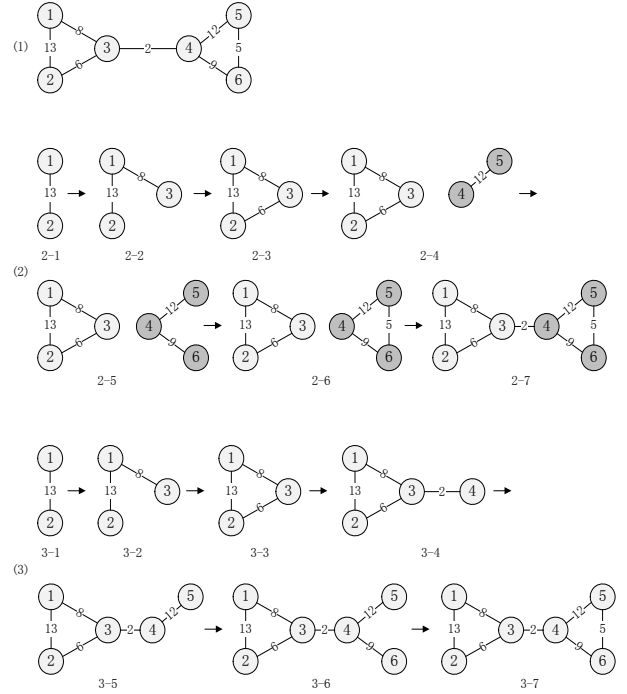

**Figure 2. The evolution of a dynamic network.**

From this example it is observed that different orders of increased edges result in completely different community structures, the first one of which is consistent with the result of the BGL algorithm. This is mainly because the community structure is built from null and there is no heuristic information about the initial partition of the network. If the network is large and there is a good initial partition, the above situation is hardly to happen, especially when the edges are randomly increased, which we suppose to be the case of most real world networks. An important reason we choose the BGL algorithm to generate the initial community structure is because its quality is quite high, compared to other algorithms.

## 4.4 Complexity Analysis

Since our incremental algorithm orients from the BGL algorithm, we will first review the computational complexity of it. The BGL algorithm is iterative and we call each round of iteration as a *pass*, which is further divided into two phases. Assume that we start with a weighted network of $N$ nodes and $M$ edges. In the first phase, each node is considered to be an isolate community. So there are as many communities as number of nodes in this initial partition. For each node $i$ we compute the gain of modularity by moving $i$ from its current community to the neighbor communities. Then $i$ is moved to the community for which the gain of modularity is the maximum and positive, otherwise $i$ will stay in its original community. From Eq. (4) we see the time complexity for moving node $i$ to its neighbor community $c_j$ is $O(1)$. For node $i$, all of its neighbor nodes will be considered to get the maximum grain of modularity. So the computing time for moving node $i$ is $O(d_i)$, where $d_i$ is the degree of node $i$. In the first phase each node has to be handled, so the overall time complexity of the first phase is $\sum_{i=1}^{N} O(d_i) = O(M)$. After the first phase many isolate communities will disappear while some of them will become larger, making the total number of communities much smaller. The second phase of the algorithm consists in building a new network whose nodes be the communities found during the first phase. The main work during this phase is summing up the weights of edges between nodes of communities. In practical implementation, this work has been done during the first phase, which means the total computing time of the first pass is $O(M)$. Assume that after the pass we get a network of $N_1$ nodes and $M_1$ edges, then the time complexity of the second pass will be $O(M_1)$. If the algorithm stops after the $k$th pass, the total time complexity of this algorithm will be $O(M + \sum_{i=1}^{k-1} M_k)$. The experimental results show that the number of passes is usually smaller than 5 and the algorithm converges very quickly after the first pass, due to the quick decrement of number of edges and nodes after a pass. So the computing time is mainly spent on the first pass, making the total time complexity of the algorithm the same magnitude as $O(M)$.

In the incremental algorithm, different types of increased edges correspond to different operations, therefore with different computing time. For an increased edge $l(i,j,w_{ij})$, if operation 1 is taken, since no new node is increased and all nodes keep their community affiliation unchanged, all that needs to do is updating $\Sigma_{in}^i$ and $\Sigma_{tot}^i$ with $O(1)$ time. Similarly, the time complexity of operation 3 and operation 4 is $O(1)$. For operation 2, the two communities $c_i$ and $c_j$ are combined into one. So each vertex in one of the communities (assume $c_j$) has to change its community affiliation to the other. Suppose the size of the new community is $S$, then the time complexity will be $O(S)$. According to the statistical results of our experiments, the percentage of occurrences of operation 2 is usually very low (often less than 1%) compared to that of other operations, so the incremental algorithm will run at a high speed in most of the time.

## 5. PERFORMANCE EVALUATIONS

In this section we will evaluate the performance of our incremental algorithm by the modularity on four real world datasets, including communication networks, citation networks, and social networks. In our experiments we model the change of network as sequential increment of edges, while not consider the decrease of edges. This is because for real word networks, we usually know from the observation of interactions of two individuals that a relationship is built, while we could not tell that the relationship is ended by the same observation. The experimental results are compared with the well known CNM and BGL algorithms. We also compare the computing time of our algorithm with that of CNM and GBL when dealing with dynamic networks. It is shown that our algorithm outperforms both BGL and CNM in computing time and outperforms CNM in terms of modularity.

### 5.1 Evaluations Under the Enron Dataset

The Enron dataset [28] collects email communication documents from 150 senior executives in the Enron corpus. This dataset was originally prepared by the CALO Project [1] (A Cognitive Assistant that Learns and Organizes) and contains a total of about 0.5M messages. Then a number of folks and organizations worked hard on it to correct the problems (e.g. invalid email addresses, etc.) of the dataset, making it more comfortable to researchers. By now this dataset has been largely used by researchers who are interested in community discovery, dynamic social networks, email tools improvement, etc. The latest version of the dataset (about 423M) was published in August 21, 2009. The elder version (March 2, 2004) is no longer accessible.

The dataset consists of 36,692 vertexes and 367662 edges. We get the network data from the web page of Stanford SNAP Graph Library [2], who has preprocessed the communication data and generates a well organized text file. The first several lines of the text file give some simple introduction about the dataset, after that are $M$ lines (edges), each of which is a pair of integer numbers identifying the two vertexes of the edge. The network is directed and unweighted, so the weight of each edge will be assigned 1. Since our algorithm deals with undirected network, edges $l(i,j,1)$ and $l(j,i,1)$ will be treated as a single edge. So the number of edges is reduced to 183,831.

#### 5.1.1 Modularity

We firstly evaluate the quality of community structure by its *modularity*. The network data (edges) is randomly divided into two parts with equal size: the original data and the incremental data. For the original data, we perform the BGL algorithm to generate an initial partition of the community structure. The CNM algorithm is also performed as a comparison. For the incremental data, we add the edges to the network sequentially to update the community structure. Due to the high computing complexity of the BGL and CNM algorithm and the large amount of incremental data, it is not possible to finish the experiment within finite time (24 hours) if they are performed after each edge is increased. So we further divide the incremental data into ten equal subsets, which means that each subset contains 5% of the data. To update the community structure, the incremental algorithm is performed after each edge is increased while the BGL and CNM algorithm is performed on the aggregated data only after all the edges of a subset are increased. After each subset is increased, we calculate the *modularity* of the community structure. Figure 3 shows how the modularity changes along with the increase of subsets.

We see that the modularity value of the BGL algorithm (denoted with triangles) varies obviously among different executions, from the lowest 0.5935 to the highest 0.6346. It is because the algorithm introduces a random mechanism to improve its performance. Since the incremental algorithm take advantage of the BGL algorithm to generate the initial partition, they have the same modularity value in the beginning. After the subsets are sequentially added, the modularity of the incremental algorithm gradually decreases, from 0.6319 to 0.5925, about 6% decrement. Even so, the smallest modularity value is much better than that of the CNM algorithm, about 0.5106. Also, we should observe this decrement is produced after half of the total data is increased. In applications we can perform the BGL algorithm periodically to generate static community structure at different time steps and then apply our incremental algorithm to track the dynamic community structure. It is noteworthy that the modularity decreases very slowly. For example, if we take the first subset as the overall incremental data, the modularity will decrease from 0.6319 to 0.6262, about only 0.9% decrement. If the initial partition of network is good enough, it is believed that the result will be even better.

---

[1] http://www.ai.sri.com/project/CALO

[2] http://snap.stanford.edu/data/

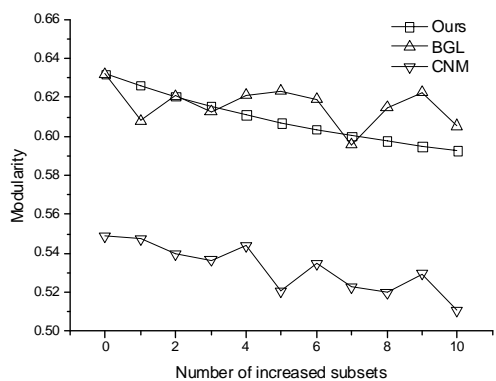

**Figure 3. The change of modularity over the increment of subsets**

### 5.1.2 Time Complexity

Compared to its promising behavior in modularity, the biggest advantage of our incremental algorithm is its high time efficiency, which makes it possible to track the community structure of networks in a fine-grained way. We compare the overall computing time spent on tracking community structure by our algorithm with that of CNM and BGL. Each of the ten subsets of incremental data includes about 9,000 edges. Once an edge is increased, the incremental algorithm is performed to update the community structure. If all the 9,000 edges of a subset are increased, the BGL and CNM algorithm will be performed as comparisons. The result is shown in Figure 4. The curves denote the relationship between the overall computing time and the number of increased subsets. We see that our incremental algorithm only takes 0.205 seconds to track the community structure of the network whose amount of edges varies from about 90,000 to 183,831, while the BGL and CNM algorithms needs 24 and 712 seconds respectively. Even so, they only get ten snapshots of the community structure, still unable to keep track of it on a fine-grained level.

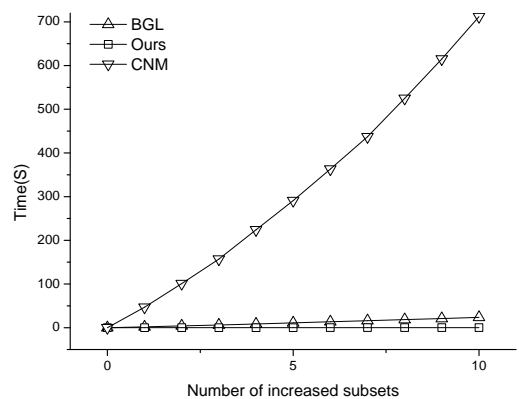

**Figure 4. Computing time over the increment of subsets.**

## 5.2 Other Real World Datasets

We also test our incremental algorithm on three other real world datasets: the Arxiv High Energy Physics Theory paper citation network (cit-HepTh), the Wikipedia who-votes-on-whom network (wiki-Vote) and the ForSun corp IM communication network (FSIM). Table 2 gives a simple summarization of these three datasets.

**Table 2. Introduction to the three real world datasets**

| Name | Nodes | Edges | Type |
|---|---|---|---|
| cit-HepTh | 27,770 | 352,807 | unweighted, directed |
| wiki-Vote | 7,115 | 103,689 | unweighted, directed |
| FSIM | 14,472 | 12,120 | weighted, undirected |

The Arxiv HEP-TH citation network comes from the e-print arXiv[3] and covers all the citations within a dataset of 27,770 papers with 352,807 edges. If a paper $i$ cites paper $j$, there is a directed edge in the network graph points from $i$ to $j$. If a paper cites, or is cited by a paper outside the dataset, no edge will be created. The data of wiki-Vote is from social community Wikipedia[4], a free encyclopedia written collaboratively by volunteers around the world. Some of the Wikipedia contributors are administrators, who are users with access to additional technical features. If a user wants to become an administrator, he has to be issued a request for adminship and then voted by the community. The Stanford SNAP graph library[2] extracted all the administrator elections and vote history data (till January 3 2008), resulting in a network graph, where vertexes represent Wikipedia users and a directed edge from vertex $i$ to $j$ represents that user $i$ voted on user $j$. We get the two network graphs from the Stanford SNAP graph library and convert them into weighted (each edge is assigned with weight 1) undirected graphs. The cit-HepTh and wiki-Vote are both well known datasets, different from them, the FSIM dataset is provided by a company named ForSun corp. The network graph is modeled by the instant message communication data of about 2,000 volunteers. For privacy consideration their personal information will not be published. The data is collected from July 2011 to October 2011, about 3 months of data, giving us a network graph of 14,472 vertexes and 12,120 edges. In the network each vertexes represent a user, an undirected edge from vertex $i$ to $j$ represents that there are messages exchanged between the two users. The weight of an edge is evaluated by the frequency of the communication.

The experimental results of the three datasets are show in Figure 5, where we see our incremental algorithm outperforms that of CNM's on both modularity and computing time in the first two datasets. In Figure 5-(c) we find an interesting thing that the BGL and CNM algorithms almost get the same modularity value. This is mainly caused by the special network structure of the dataset. In fact, the network graph is not a connected graph and has many isolated star-like communities. An isolated star-like community is intuitively defined to contain a central node and several neighbors connecting to the center, and there are no connections between the nodes in the community to the rest of the graph. The best performance of our algorithm is on the wiki-Vote dataset. When all the subsets are increased to the network, the decrements of modularity for the three datasets are 3.9%, 3.71% and 1.06% separately, which are acceptable.

[3] http://arxiv.org/

[4] http://www.wikipedia.org/

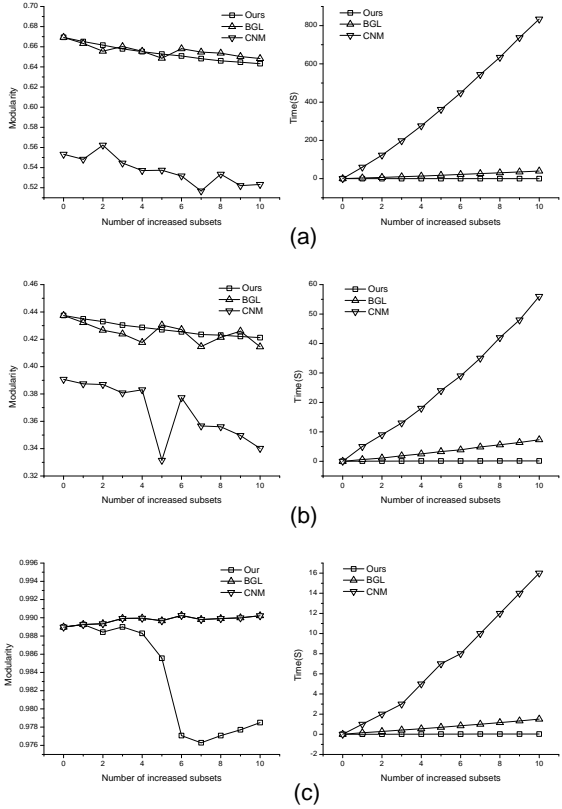

**Figure 5. Experimental results on three other real world datasets: (a) cit-HepTh; (b) wiki-Vote; (c)FSIM**

As we mentioned in section 3, different types of edges are handled by different operations, which correspond to different time complexity. Figure 6 gives the statistical information about the percentage of operations taken in updating the community structure with our incremental algorithm. The percentage of operation 2 on the four datasets is 0.406%, 0.113%, 0.014% and 2.8% (three of which are lower than 1%), which guarantees the high time efficiency of our algorithm. We observe an interesting thing that the operation distribution on the FSIM dataset is quite different from the former three datasets. We guess this may be caused by the special sparsity and star-like characteristics of the dataset. This may also explain why the modularity value of our algorithm is lower than that of CNM on the FSIM dataset.

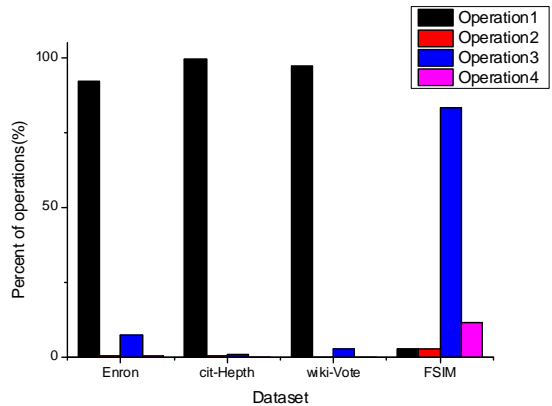

**Figure 6. Distribution of operations taken in updating the community structure**

In table 3 we summarize the modularity value and computing time of our algorithm and that of BGL and CNM on the four datasets after all subsets are increased.

**Table 3. Modularity and computing time**

| Q/Time(s) | CNM | BGL | Our |
| --- | --- | --- | --- |
| **Enron** | 0.5106/712 | 0.6053/24 | 0.5926/0.2 |
| **cit-HepTh** | 0.5233/834 | 0.6483/40 | 0.6432/0.23 |
| **wiki-Vote** | 0.3402/56 | 0.4125/7.3 | 0.4212/0.113 |
| **FSIM** | 0.9902/16 | 0.9902/1.5 | 0.9785/0.03 |

## 6. CONCLUSION

In this paper we propose an incremental algorithm to keep track of community structure of dynamic networks which is large and change frequently. The community structure is defined as divisions of network vertexes into sub groups, within which nodes are densely connected while between which they are sparsely connected. In 2004, Newman proposed a metric called *modularity* to evaluate the quality of a community structure. Though finding the optimal solution is NP hard, a lot of work is done to find the suboptimal solutions, and the performance of some proposed algorithms is quite good. However, most of the modularity optimization based algorithms are static and just give a snapshot of the networks. When dealing with tracking community structure which changes from time to time, they all suffer from high computing complexity. Many researchers propose algorithms to deal with dynamic networks, but they are still unable to track the community structure on a fine-grained level. The greatest strength of our incremental algorithm is its low computing complexity, as low as $O(1)$, making it possible to keep track of the community structure of dynamic networks in a fine-grained way. The efficiency of our algorithm is evaluated by both modularity and its computing time. We compare our algorithm with the ones proposed in [15] and [21]. The experimental results show that our algorithm has reasonably good performance on both modularity and computing time.

As we discussed in section 3, the performance of our incremental algorithm partly depends on the initial community structure of the network. However, finding a "good" initial community structure is not an easy thing. In this paper we use the *modularity* as the metric to evaluate the community structure. But high modularity does not necessarily correspond to good community structure. In the future we will try some other metrics to evaluate our incremental algorithm. In the algorithm we model the change of networks as sequential increment of edges, while not consider the decrement of edges. This will also be considered in our future work.

## 7. ACKNOWLEDGMENTS

We thank Anyan Chen, Lei Xu, Handong Mao, Chengkai Guo, Gang Wang, Lei Guo and Xilong Jin for their help in finishing this work. Special thanks to Deke Guo and Yingwen Chen, they provide us with much precious advice.